\documentclass{ws-procs975x65}
\def\beq{\begin{equation}}
\def\eeq{\end{equation}}
\begin{document}
\title{Commissioning of ALFABURST: initial tests and results}
\author{Kaustubh Rajwade$^1*$ , Jayanth Chennamangalam$^2$, Duncan Lorimer$^1$, Aris Karastergiou$^2$, Dan Werthimer$^3$, Andrew Siemion$^3$, David MacMahon$^3$, Jeff Cobb$^3$, Christopher Williams$^2$ and Wes Armour$^4$}

\address{$^1$Department of Physics and Astronomy, West Virginia University,\\
Morgantown, WV 26506, USA\\
$^*$E-mail: kmrajwade@mix.wvu.edu\\
}
\address{$^2$Astrophysics, University of Oxford, Denys Wilkinson building,\\
Keble Road, Oxford OX1 3RH, United Kingdom\\
}
\address{$^3$Department of Astronomy, University of California Berkeley,\\
Berkeley, CA 94720, USA\\
}
\address{$^4$Oxford e-Research Centre, University of Oxford, Keble Road,\\
Oxford OX1 3QG, United Kingdom\\}

%\author{Jayanth Chennamangalam, Aris Karastergiou, Christopher Williams and Wes Armour$^1$} 

%\address{Astrophysics, University of Oxford, Denys Wilkinson building,\\
%Keble Road, Oxford OX1 3RH, United Kingdom\\
%$^1$Oxford e-Research Centre, University of Oxford, Keble Road,\\
% Oxford OX1 3QG,United Kingdom\\
%}

%\author{Dan Werthimer, Andrew Siemion, David MacMahon, Jeff Cobb}
%\address{Space Science Lab, University of California, Berkeley,\\
%Berkeley, CA, USA\\
%}

\begin{abstract}
Fast Radio Bursts (FRBs) are apparently one-time, relatively bright radio 
pulses that have been observed in recent years. The origin of 
FRBs is currently unknown and many instruments are being built to detect more 
of these bursts to better characterize their physical properties and identify 
the source population. ALFABURST is one such instrument. ALFABURST takes 
advantage of  the 7-beam Arecibo L-band Feed Array (ALFA) receiver on the 
305-m Arecibo Radio Telescope in Puerto Rico, to detect FRBs in real-time at 
L-band (1.4 GHz). We present the results of recent on-sky tests and 
observations undertaken during the commissioning phase of the instrument. 
ALFABURST is now available for commensal observations with other ALFA projects.
\end{abstract}

\bodymatter

\section{Introduction}
FRBs have been a mystery to astronomers since their discovery in 2007 ~\cite{drl07}. Since then, the number of confirmed detections has gone up 
to $\sim$15~\cite{pet15,th13,spi14,Ch15}. The origin of these bursts are still 
unknown. The dearth of discovery of these bursts has resulted in development 
of instruments at different radio telescopes capable of detecting short, 
sporadic radio pulses. Recent efforts in time-domain radio astronomy have 
focused on real-time FRB detection with the promise of rapid follow-up of new 
events. Here we present the results from the recent commissioning phase of one 
such instrument, ALFABURST which has been developed for the Arecibo Radio 
telescope to monitor the sky for FRBs in real time.
\section{Commissioning phase}
\subsection{Bandpass validation}
\begin{figure*}
%\centering{\epsfig{file=0611+all_beams.eps,scale=0.2}}
\includegraphics[width=130mm,height=70mm]{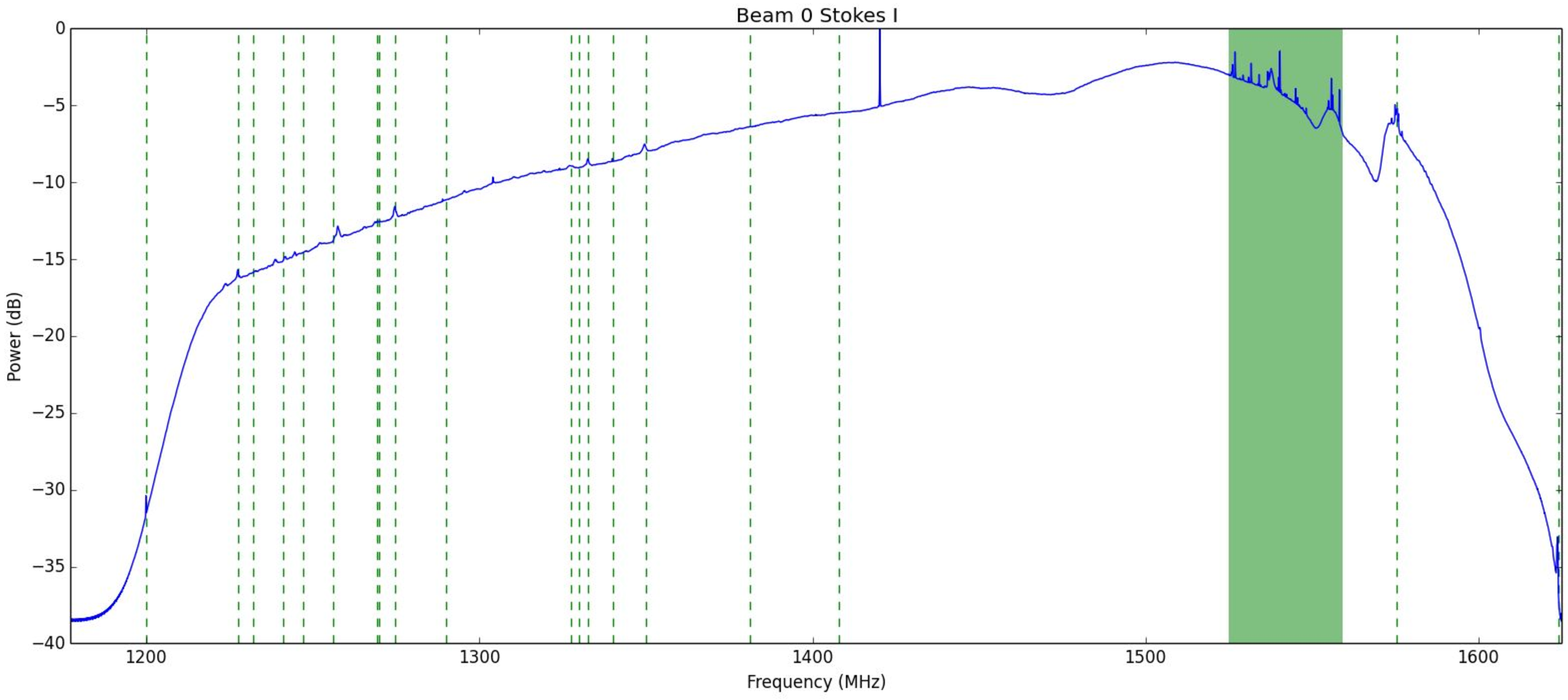}
\caption{Bandpass obtained from 1 of the 7 beams at L-band from a 5~minute observation using ALFABURST. The dashed vertical lines and the shaded region represent the known sources of RFI.
\label{fig:band}}
\end{figure*}

During the commissioning phase of the instrument, a number of validation tests were performed. One of them included an on-sky test to collect data using the instrument and generate a bandpass. The ALFABURST hardware acquired the data from which the bandpass was obtained. The bandpass was also scrutinized to identify known sources of radio frequency interference (RFI). The peaks in the bandpass matched with the known RFI sources confirming that we obtained the correct band from the instrument (see Figure~\ref{fig:band}).

\subsection{Test observations}
\begin{figure*}
%\centering{\epsfig{file=0611_all_beams.eps,scale=0.2}}
\includegraphics[width=140mm,height=80mm]{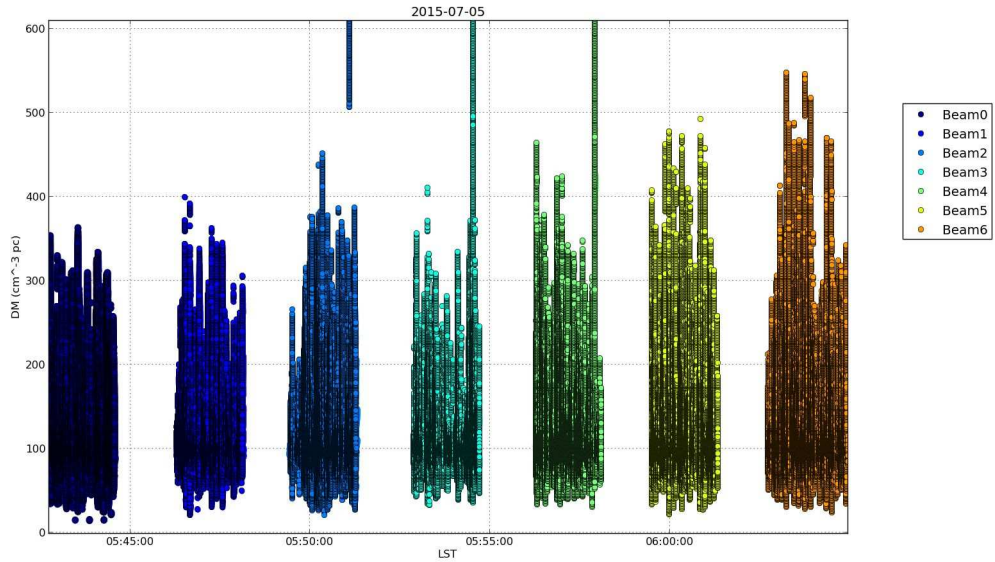}
\caption{DM versus time plot of detection of PSR B0611+22 in all the seven beams of the ALFA receiver.
\label{fig:0611}}
\end{figure*}

After the initial tests were successful, we carried out observations with the 
instrument with dedicated time on the telescope. A few candidates (pulsars 
which emit bright pulses) were selected for the observation. ALFABURST looks 
for bright pulses in real-time using the ARTEMIS software pipeline~\cite{ak15}. We observed PSR~B0611+22 and PSR~B0531+21 (Crab Pulsar) in each of the beams of the ALFA receiver for a span 
of $\sim 2$~minutes. Fig.~\ref{fig:0611} shows the detection of bright pulses 
for pulsar PSR B0611+22 in all the seven beams of ALFA. The pulses are detected at the dispersion measure (DM) of PSR~B0611+22 ($\sim 97 \rm cm^{-3}~\rm pc$ ). 
These observations were also beneficial for testing the beam to backend mapping of the instrument to make sure that we are detecting the pulsar in the expected data stream.
We also tested the system in commensal mode to make sure that we're able to start the pipeline when ALFA is enabled and stop it when it's disabled.

\section{Conclusions and future work}
After successfully testing the instrument, ALFABURST is online to 
detect FRBs in real-time at Arecibo . Future work includes doing more 
observations to verify if the whole pipeline is working smoothly over a wider 
bandwidth. Currently, we are able to process data in real-time over a 
bandwidth of 56 MHz with a DM limit of 2560~$\rm{cm^{-3}~\rm pc}$. Eventually, we plan on operating ALFABURST at the full bandwidth of 300 MHz over a maximum DM of $\sim$10,000~$\rm{cm^{-3}~\rm pc}$.

%\bibliography{proc}{}
%\bibliographystyle{ws-procs975x65}
\end{document}